\documentclass[pra,twocolumn,showpacs,groupedaddress,nofootinbib]{revtex4-1}
\usepackage{amsmath,amssymb,amsfonts}
\usepackage{graphicx}
\usepackage{txfonts}
\usepackage{setspace}

\begin{document}

\title{Bose-Einstein condensation of $^{85}$Rb by direct evaporation in an optical dipole trap}

\author{A. L. Marchant}
\affiliation{Department of Physics, Durham University, Durham DH1 3LE, United Kingdom}
\author{S. H\"{a}ndel}
\affiliation{Department of Physics, Durham University, Durham DH1 3LE, United Kingdom}
\author{S. A. Hopkins}
\affiliation{Department of Physics, Durham University, Durham DH1 3LE, United Kingdom}
\author{T. P. Wiles}
\affiliation{Department of Physics, Durham University, Durham DH1 3LE, United Kingdom}
\author{S. L. Cornish}
\affiliation{Department of Physics, Durham University, Durham DH1 3LE, United Kingdom}

\date{\today}

\begin{abstract}
We report a simple method for the creation of Bose-Einstein condensates of $^{85}$Rb by direct evaporation in a crossed optical dipole trap. The independent control of the trap frequencies and magnetic bias field afforded by the trapping scheme permits full control of the trapped atomic sample, enabling the collision parameters to be easily manipulated to achieve efficient evaporation in the vicinity of the 155~G Feshbach resonance. We produce nearly pure condensates of up to $4\times10^4$ atoms and demonstrate the tunable nature of the atomic interactions.
\end{abstract}

\pacs{
03.75.Nt,    
37.10.De,    
67.85.Hj,    
}

\maketitle

\section{Introduction}
The use of magnetically tunable Feshbach resonances \cite{PhysRevA.47.4114} to control the interaction between atoms is now commonplace in many ultracold atomic gas experiments. The ability to precisely tune the $s$-wave scattering length, $a$, near a broad resonance has resulted in many exciting breakthroughs in the study of atomic Bose-Einstein condensates (BECs) and degenerate Fermi gases \cite{RevModPhys.82.1225, RevModPhys.80.1215}. At the same time, narrow resonances have found applications in the coherent association of ultracold molecules \cite{RevModPhys.82.1225, RevModPhys.78.1311}, bringing quantum degenerate samples of ground state molecules within reach \cite{Danzl.rovib, Ni10102008}. Although essentially all single-species alkali-metal atoms exhibit some sort of Feshbach spectrum, broad resonances suitable for tuning the scattering length are generally less accessible. For example,  the broadest resonance in $^{87}$Rb, the workhorse of many quantum gas experiments, is at a field of 1007~G and just 0.2~G wide \cite{PhysRevA.68.010702}. In contrast, a resonance exists for $^{85}$Rb atoms in the $F = 2$, $m_F = -2$ state at 155~G which is 10.7~G wide \cite{PhysRevLett.81.5109, PhysRevA.67.060701} yielding a variation of the scattering length with magnetic field of $\sim40~a_0$~G$^{-1}$ in the vicinity of $a=0$. This has already been used successfully to precisely tune the atomic interactions in a BEC \cite{PhysRevLett.85.1795, PhysRevLett.86.4211, Nature.Collapse, PhysRevLett.96.170401, PhysRevLett.101.135301, PhysRevA.84.033632}. Despite this, $^{85}$Rb has been notably underused in quantum gas experiments owing to its perceived reputation as a difficult species to cool to degeneracy.

\begin{figure}
	\centering
		\includegraphics{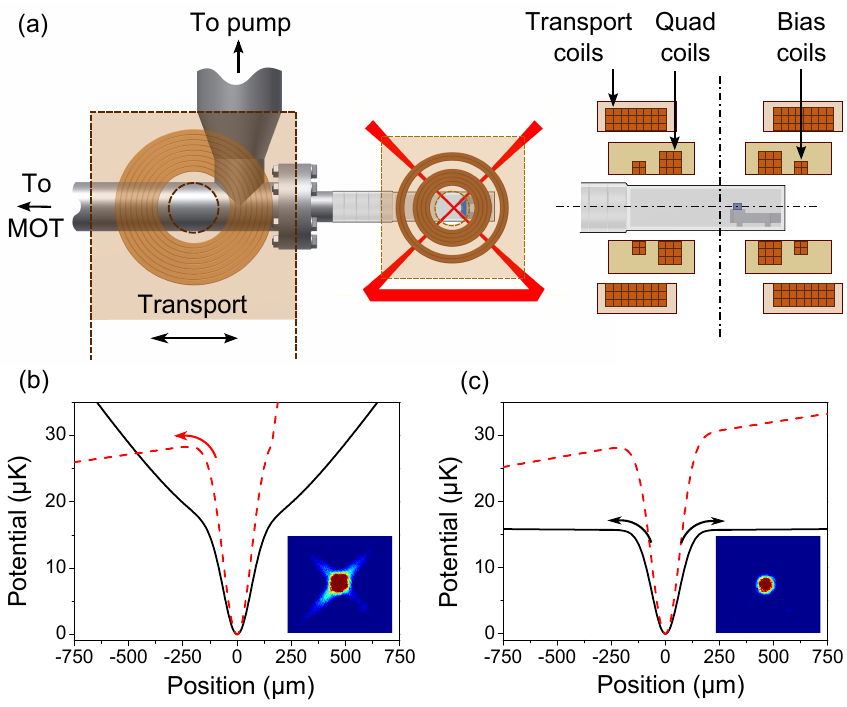}
	\caption{(Color online) (a) Experimental setup showing the arrangement of coils around the UHV cell and the beam geometry used to create the crossed dipole trap. Trapping: Potentials produced horizontally along one of the beams (black, solid) and vertically (red, dashed) in the hybrid (b) and levitated (c) crossed dipole traps. Insets: False colour images of atoms in the respective traps, viewed from above. }
	\label{Fig1:setup}
\end{figure} 

The difficulties associated with attempting to evaporatively cool $^{85}$Rb are well documented \cite{PhysRevLett.85.1795, PhysRevLett.80.2097, PhysRevLett.85.728}. The elastic collision rate in samples trapped directly from a magneto-optical trap (MOT) is severely suppressed due to an unfortunately placed zero in the $s$-wave scattering cross section \cite{PhysRevLett.80.2097}. Additionally, the two- and three-body inelastic collision rates in ultracold samples are unusually high and vary strongly in the vicinity of the Feshbach resonance \cite{PhysRevLett.85.728}. Nevertheless, carefully optimized evaporation in a weak Ioffe-Pritchard magnetic trap produced stable condensates of $\sim10^4$ atoms \cite{PhysRevLett.85.1795}. However, with the development of optical trapping, the modern evaporator is equipped with a broader array of tools than her predecessor, allowing her to navigate the potential pitfalls associated with $^{85}$Rb with greater ease. For example, recent experiments \cite{PhysRevLett.101.040402, Altin.RevSciInst} have almost circumvented these problems entirely by employing $^{87}$Rb to sympathetically cool low density samples of $^{85}$Rb, yielding condensates of up to $8\times10^4$ atoms \cite{PhysRevLett.101.040402}, at the expense of added experimental complexity. 

Here we report a simple method for the creation of Bose-Einstein condensates of  $^{85}$Rb by direct evaporation in a crossed optical dipole trap. The independent control of the trap frequencies and magnetic bias field afforded by the trapping scheme permits full control of the trapped atomic sample, enabling the collision parameters to be easily manipulated to achieve efficient evaporation in the vicinity of the 155~G Feshbach resonance. We produce nearly pure condensates of up to $4\times10^4$ atoms and demonstrate the tunable nature of the atomic interactions. Finally we briefly discuss the application to future work on condensate collapse and soliton formation.

The structure of the paper is as follows. In section \ref{expt} we describe the experimental setup, detailing the loading of the crossed dipole trap. In section \ref{evap} we outline the evaporation scheme used cool the atomic sample and address the experimental complexity associated with the cooling of $^{85}$Rb. Section \ref{BEC} details the final parameters used to reach BEC. In section \ref{tune} we demonstrate the tunable nature of the atomic interactions in the condensate before concluding and presenting our plans for future work in section \ref{conclusion}.

\section{Experimental Overview}
\label{expt}
Central to our setup is a crossed dipole trap which we produce inside a UHV glass cell, shown in Fig.~\ref{Fig1:setup}(a). Atoms are delivered to the cell via a magnetic transport apparatus, details of which can be found in \cite{SylviApparatus}. The final stage of the delivery process is to transfer atoms into a static quadrupole trap (axial gradient  180~Gcm$^{-1}$) constructed around the glass cell. The transport trap is then moved away from the glass cell to allow greater optical access. Typically the trapped atomic cloud contains $\sim5\times$10$^8$ atoms at a temperature of $\sim380~\mu$K. Forced RF evaporation of the sample is then carried out. The optimized evaporation trajectory takes 26~s in total due to relatively slow rethermalisation caused by the low $s$-wave scattering cross section in this temperature range \cite{PhysRevLett.80.2097}. However, we note that the linear potential produced by the quadrupole trap means it is possible to obtain runaway evaporation for a lower ratio of elastic to inelastic collision compared to a harmonic potential \cite{KetterleReview}. The RF evaporation results in a cloud of $3\times10^7$ atoms at 42~$\mu$K with a phase space density (\textit{PSD}) of $5\times10^{-5}$. At this temperature the quadrupole trap lifetime is limited by Majorana spin flips as the coldest atoms spend increasing amounts of time close to the magnetic field zero. To proceed, we transfer a fraction ($\sim20~\%$) of the atomic sample into the crossed optical dipole trap and, in doing so, gain a factor of 30 in \textit{PSD}. 

\begin{figure}
	\centering
		\includegraphics{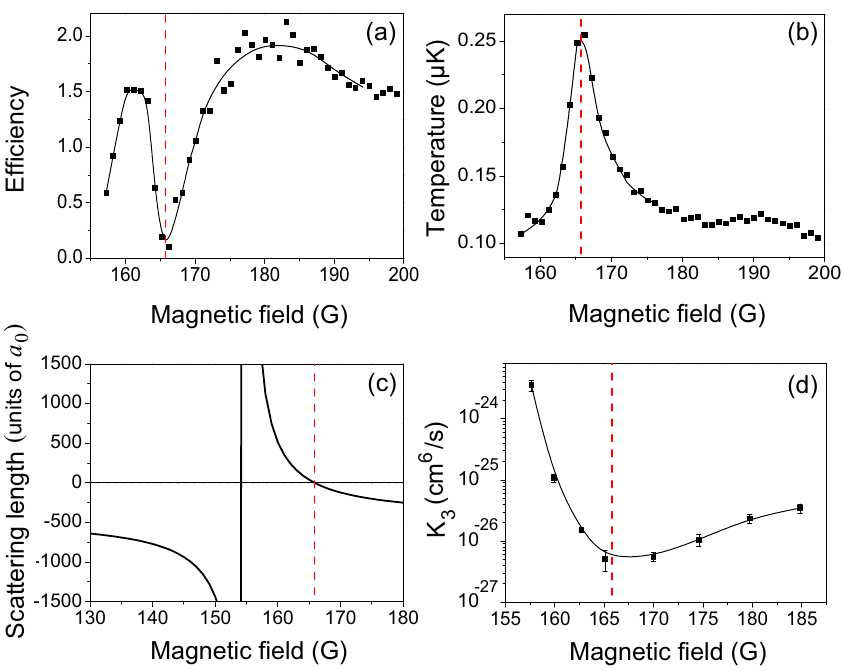}
	\caption{(Color online) Collisional properties. Position of $a=0$ is marked by the red, dashed line: (a) Evaporation efficiency (see text) of a fixed evaporation sequence carried out at different magnetic fields close to the Feshbach resonance. (b) The effect on fitted temperature of the same evaporation ramps. (c) Feshbach resonance in the $F=2, m_F=-2$ state of $^{85}$Rb. The scattering length is given in units of the Bohr radius, $a_0\approx 0.529 \times 10^{-10}$~m. (d) Magnetic field dependence of the three-body inelastic loss rate near the Feshbach resonance. Solid lines in (a), (b) and (d) are a guide to the eye only. }
	\label{fig:K3Loss}
\end{figure}

The crossed dipole trap is produced using a bow-tie arrangement of a single $\lambda=1064$~nm beam with a total incident power of 10.1~W derived from a single frequency fiber laser (IPG: YLR-15-1064-LP-SF). The beam is initially focussed to $\sim136~\mu$m. It then exits the cell and is refocussed to produce an effective second beam of waist $\sim125~\mu$m, crossing the first beam at 90$^\circ$. The smaller waist compensates for reflection losses at the glass cell windows. The beams intersect $\sim160~\mu$m below the quadrupole field zero to produce trap frequencies of $\omega_{x,y,z}=2 \pi \times$(190, 160, 250)~Hz at full power. Here $x$ and $y$ are along the dipole beams and $z$ is in the vertical direction. The dipole trap is switched on during the RF evaporation stage. To complete the loading, the quadrupole gradient is relaxed from 180~Gcm$^{-1}$ to $\sim21.5$~Gcm$^{-1}$ in 500~ms. The final quadrupole gradient used is just less than that sufficient to support atoms against gravity ($\sim22.4$~Gcm$^{-1}$). The presence of the magnetic gradient leads to weak magnetic confinement along the beams \cite{PhysRevA.79.063631}, enhancing the trap volume and resulting in a trap depth set by the vertical direction (Fig.~\ref{Fig1:setup}(b)). In addition, the offset from the field zero leads to a small magnetic field ($\sim0.3$~G) at the location of the crossed dipole trap. As the atoms are subject to both optical and magnetic confinement we refer to this as the hybrid trap. 

\section{Evaporative cooling}
\label{evap}

In order to reach the Feshbach resonance it is necessary to apply a moderate bias field ($\sim155$~G). This leaves the trapped atoms still levitated against gravity however, the confinement produced by the quadrupole trap along the dipole beams is effectively removed. Atoms now only remain at the intersection of the trap. This is our levitated crossed optical dipole trap. Before switching on the bias field, we first evaporate by ramping down the power of the hybrid trap at low field where inelastic losses are known to be low \cite{PJprivate}. This allows us to exploit the enhanced volume of the hybrid trap to improve the transfer into the levitated crossed dipole trap. In contrast to the hybrid case, in the levitated trap evaporation happens preferentially along the dipole beams, as shown in Fig.~\ref{Fig1:setup}(c), resulting in a trap with roughly half the depth compared to the zero field case.

To achieve efficient evaporation in the levitated trap it is necessary to understand the interplay between elastic and inelastic collisions close to the Feshbach resonance. We explore this by indirectly probing the collision ratio, carrying out a fixed evaporation sequence for different magnetic fields. The efficiency, $\gamma$, of the evaporation sequence can then be calculated from the initial (\textit{i}) and final (\textit{f}) number, \textit{N}, and \textit{PSD} of the gas according to: 
\begin{equation*}
\gamma=-\frac{\log(\textit{PSD}_f/\textit{PSD}_i)}{\log(N_f/N_i)}.
\end{equation*}  
The efficiency for a 50~G window spanning the zero crossing of the Feshbach resonance is shown in Fig.~\ref{fig:K3Loss}(a). The two clear peaks at 161~G and $175-185$~G highlight the most efficient fields at which to evaporate with the broad $a<0$ peak, $175-185$~G, giving marginally better performance. As $a$ approaches zero (red dashed line) the elastic collision rate reduces and rethermalisation ceases. As a result the efficiency tends to zero and we see a corresponding peak in the fitted cloud temperature, Fig.~\ref{fig:K3Loss}(b).  

The distinct structure evident in Fig.~\ref{fig:K3Loss}(a) follows from the magnetic field dependence of the elastic and inelastic collision rates. The elastic collision rate, determined by the atomic scattering length shown in Fig.~\ref{fig:K3Loss}(c), varies by many orders of magnitude over the region of interest. Similarly, inelastic losses are known to exhibit a strong field dependence close to the resonance. Fig.~\ref{fig:K3Loss}(d) shows the change in the three-body inelastic loss rate, $K_3$, measured close to the Feshbach resonance \footnote{An atomic sample at $\sim0.15~\mu$K is produced at 175~G. The magnetic field is then ramped from 175~G to a new value in 10~ms. Following the ramp, the trap is compressed (in 1~s) by linearly increasing the dipole beam power. This deepens the trap from 1.4~$\mu$K to 27.5~$\mu$K, increasing the atomic density and putting us well into the three-body dominated loss regime. The lifetime of the atomic cloud in this new trap is measured and the three-body loss rate, $K_3$, determined from a fit to the atom number evolution with time \cite{PhysRevLett.91.123201}.}. It is apparent that for a given magnitude of scattering length the inelastic losses are marginally lower on the $a<0$ side of the zero crossing as previously predicted and observed \cite{PhysRevLett.83.1751, PhysRevLett.85.728}. This leads to slightly better evaporation performance for the $175-185$~G peak (Fig.~\ref{fig:K3Loss}(a)).

\section{Bose-Einstein condensation}
\label{BEC}

With this understanding of the collisional properties of the trapped cloud it is then clearer how best to achieve BEC in the optical trap. Following the low field evaporation stage in the hybrid trap the bias field is ramped rapidly ($\sim10$~ms) to $175-185$~G to exploit the window of efficient evaporation. It is at this low loss region that the majority of the subsequent evaporation is carried out. Following a 500~ms hold to allow atoms to equilibrate in the reduced trap depth, we apply two more evaporation ramps resulting in a sample of $2.5\times10^5$ atoms at 150~nK with a \textit{PSD} of 0.5. Unfortunately, stable condensates cannot be created at this magnetic field due to the large, negative scattering length ($\sim-205~a_0$) \cite{PhysRevA.51.4704} hence we must ramp the bias field again, this time to 161.3~G where the scattering length is positive ($\sim315~a_0$). A further evaporation ramp is carried out here reducing the beam power to 0.3~W and creating an almost spherically symmetric trapping geometry, $\omega_{x,y,z}=2 \pi \times$(31, 27, 25)~Hz. To reach BEC the sample is held in this final trap for up to 1.5~s. The total time for the evaporation sequence in the dipole trap is 14.5~s. We note that, owing to the width of the efficiency peak shown in Fig.~\ref{fig:K3Loss}(a), we are able to produce condensates of a similar size over a range of fields from 160~G to 163~G. However, below 160~G an increase in the inelastic loss rate makes condensation difficult and BECs of only a few thousand atoms are formed. 

\begin{figure}
	\centering
		\includegraphics{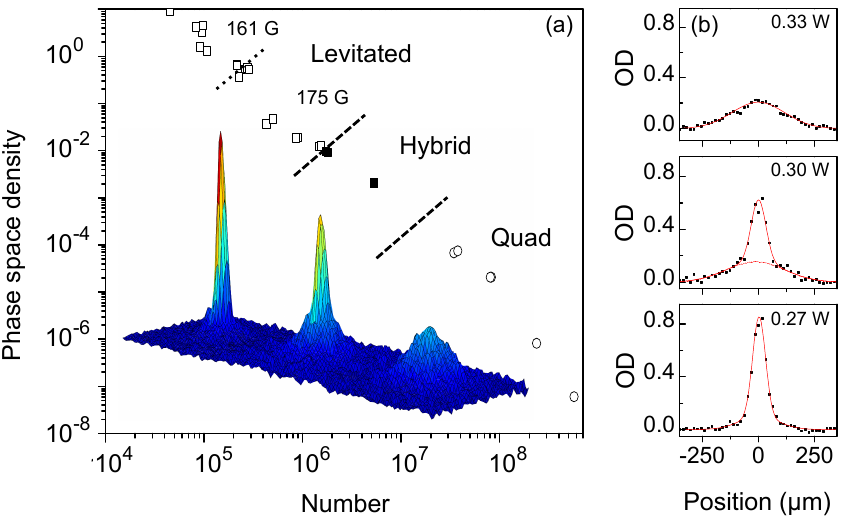}
	\caption{(Color online) Evaporation trajectory to reach BEC. (a) After RF evaporation in the quadrupole trap ($\bigcirc$), atoms are loaded into the hybrid dipole trap trap ($\blacksquare$). Following an initial evaporation stage at $\sim0$~G, a $175-185$~G bias field is applied. This produces the levitated trap ($\square$) in which further evaporation is carried out. At a \textit{PSD} of $\sim$0.5 the bias field is ramped to 161.3~G and a final evaporation stage is performed to reach BEC. Inset: Density profiles for (R-L) a thermal, bimodal and condensed atomic sample. (b) Horizontal cross-sections of the condensate column density for a thermal (top), bimodal (centre) and condensed (bottom) sample as the dipole beam power (top right) is reduced. }  
	\label{Fig:Trajectory}
\end{figure}

The complete evaporation trajectory to BEC is shown in Fig.~\ref{Fig:Trajectory}(a). Despite the difficulties associated with cooling $^{85}$Rb, it is clear that it is possible to maintain a highly efficient evaporation trajectory both in the magnetic (circles) and optical (squares) trap. Unlike the experiment in \cite{PhysRevLett.85.1795}, we do not suffer the catastrophic factor of 50 loss as we approach the BEC transition. We attribute this to a lower atomic density meaning the effect of three-body loss is not as severe in our trap. By varying the final trap depth we are able to see the transition from the thermal cloud to BEC as shown in Fig.~\ref{Fig:Trajectory}(b), with the characteristic double-distribution signature occuring with  around $10^5$ atoms in the trap. By reducing the trap depth to $\sim360$~nK we are able to produce pure condensates with $\sim4\times10^4$ atoms.

\section{Tunable interactions}
\label{tune}

\begin{figure}
	\centering
		\includegraphics{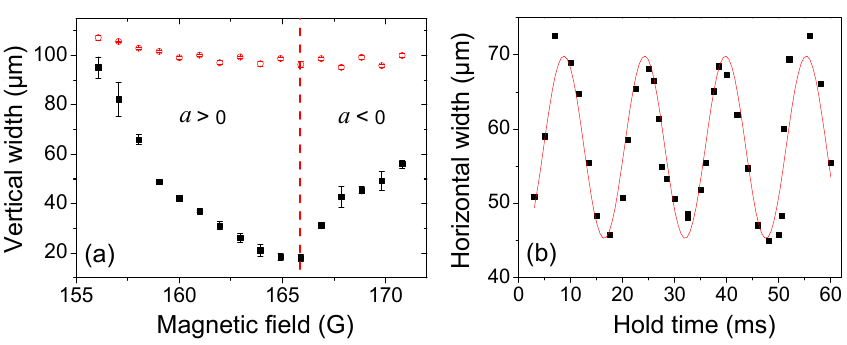}
	\caption{(Color online) Tunable interactions. (a) Change in vertical size of a pure condensate (filled squares) and thermal cloud (open circles) as a function of magnetic field applied during time of flight. (b) Breathing mode oscillation set up by jumping the atomic scattering length from $\sim315~a_0$ to $\sim50~a_0$. The hold time shown is that at the new scattering length of $\sim50~a_0$. }
	\label{fig:Tunable}
\end{figure}

In order to demonstrate the tunable nature of the condensate we present two simple experiments. In the first we alter the magnetic field synchronously with the release from the dipole trap and observe the variation of the expansion of the cloud following 55~ms of time of flight. As shown by the filled squares in Fig.~\ref{fig:Tunable}(a) the change in mean field interaction strength with magnetic field manifests itself in a change in the cloud size. We see that the BEC reaches its minimum size as the scattering length approaches zero at 165.75~G \cite{PhysRevA.64.024702}, marked by the red, dashed line. Over the region of $a>0$ the condensate number remains approximately constant. As $a$ becomes negative, the subsequent collapse \cite{Nature.Collapse} of the BEC causes an increase in cloud size. In comparison, when the same field jump is carried out using thermal atoms (for 25~ms time of flight) the comparatively low density leaves the cloud relatively insensitive to the atomic interactions and hence no change in shape is observed (open circles). The second, elegant demonstration of tunable interparticle interactions is to set up a breathing mode oscillation of the condensate by jumping the magnetic field, and hence scattering length, and observing the subsequent dynamics of the cloud at the new value of $a$. The result for a jump from $\sim315~a_0$ to $\sim50~a_0$ is shown in Fig.~\ref{fig:Tunable}(b). A jump of this type (to small $a$ but $a\neq0$) in our almost spherically symmetric trap puts us well into the Thomas-Fermi regime and hence the resulting oscillation occurs at a frequency of $\sqrt{5}\omega_{x,y}$ \cite{RevModPhys.71.463}. The trap frequency extracted using this model is in good agreement with the value measured by parametric heating.

\section{Conclusions and outlook}
\label{conclusion}

In summary, we have presented the successful cooling of $^{85}$Rb to quantum degeneracy. BEC is realised using a crossed optical dipole trap with the addition of a magnetic field gradient and a magnetic bias field to tune the atomic interactions. The atoms are cooled directly, without the need for any other refrigerant species, with pure condensates of $4\times10^4$ atoms produced at 161.3~G. Our method is very similar in spirit to a scheme first developed to cool $^{87}$Rb to degeneracy \cite{PhysRevA.79.063631} and subsequently extended to produce dual species condensates of $^{87}$Rb and $^{133}$Cs \cite{PhysRevA.84.011603}. As such, we believe our method could be readily implemented in a number of experiments currently employing $^{87}$Rb, broadening the scope for experiments with tunable interactions. However, we note that the choice of the dipole beam waists was key to the success of the trapping scheme. The high inelastic losses in $^{85}$Rb mean it is advantageous to keep trapping volumes large and trap frequencies weak. As a consequence, we found in early experiments using smaller beam waists ($\sim60~\mu$m) we were unable to reach BEC.  

In the future we plan to transfer the condensate into an optical waveguide in order to exploit the collapse shown in Fig.~\ref{fig:Tunable}(a) to produce bright matter wave solitons \cite{2002Natur.417..150S, 2002Sci...296.1290K, PhysRevLett.96.170401}. These self-stabilizing wave packets are well localized due to attractive atom-atom interactions and hence show great potential as surface probes for the study of short-range atom-surface interactions \cite{Cornish20091299}. Our apparatus includes a super-polished (surface roughness \textless1~\AA) Dove prism (shown in Fig.~\ref{Fig1:setup}(a)) for use in such experiments. In light of recent theoretical interest, there is also much scope for the study of binary soliton collisions \cite{0953-4075-41-4-045303, PhysRevA.83.041602} and the scattering of solitons from barriers \cite{PhysRevLett.102.010403, PhysRevA.80.043616} with a view to developing interferometry schemes utilizing bright matter wave solitons.

\section{Acknowledgements}
We thank C.S. Adams for many fruitful discussions. We acknowledge financial support from the UK Engineering and Physical Sciences Research Council (EPSRC grant EP/F002068/1) and the European Science Foundation within the EUROCORES Programme EuroQUASAR (EPSRC grant EP/G026602/1).



\begin{thebibliography}{36}%
\makeatletter
\providecommand \@ifxundefined [1]{%
 \@ifx{#1\undefined}
}%
\providecommand \@ifnum [1]{%
 \ifnum #1\expandafter \@firstoftwo
 \else \expandafter \@secondoftwo
 \fi
}%
\providecommand \@ifx [1]{%
 \ifx #1\expandafter \@firstoftwo
 \else \expandafter \@secondoftwo
 \fi
}%
\providecommand \natexlab [1]{#1}%
\providecommand \enquote  [1]{``#1''}%
\providecommand \bibnamefont  [1]{#1}%
\providecommand \bibfnamefont [1]{#1}%
\providecommand \citenamefont [1]{#1}%
\providecommand \href@noop [0]{\@secondoftwo}%
\providecommand \href [0]{\begingroup \@sanitize@url \@href}%
\providecommand \@href[1]{\@@startlink{#1}\@@href}%
\providecommand \@@href[1]{\endgroup#1\@@endlink}%
\providecommand \@sanitize@url [0]{\catcode `\\12\catcode `\$12\catcode
  `\&12\catcode `\#12\catcode `\^12\catcode `\_12\catcode `\%12\relax}%
\providecommand \@@startlink[1]{}%
\providecommand \@@endlink[0]{}%
\providecommand \url  [0]{\begingroup\@sanitize@url \@url }%
\providecommand \@url [1]{\endgroup\@href {#1}{\urlprefix }}%
\providecommand \urlprefix  [0]{URL }%
\providecommand \Eprint [0]{\href }%
\providecommand \doibase [0]{http://dx.doi.org/}%
\providecommand \selectlanguage [0]{\@gobble}%
\providecommand \bibinfo  [0]{\@secondoftwo}%
\providecommand \bibfield  [0]{\@secondoftwo}%
\providecommand \translation [1]{[#1]}%
\providecommand \BibitemOpen [0]{}%
\providecommand \bibitemStop [0]{}%
\providecommand \bibitemNoStop [0]{.\EOS\space}%
\providecommand \EOS [0]{\spacefactor3000\relax}%
\providecommand \BibitemShut  [1]{\csname bibitem#1\endcsname}%
\let\auto@bib@innerbib\@empty
\bibitem [{\citenamefont {Tiesinga}\ \emph {et~al.}(1993)\citenamefont
  {Tiesinga}, \citenamefont {Verhaar},\ and\ \citenamefont
  {Stoof}}]{PhysRevA.47.4114}%
  \BibitemOpen
  \bibfield  {author} {\bibinfo {author} {\bibfnamefont {E.}~\bibnamefont
  {Tiesinga}}, \bibinfo {author} {\bibfnamefont {B.~J.}\ \bibnamefont
  {Verhaar}}, \ and\ \bibinfo {author} {\bibfnamefont {H.~T.~C.}\ \bibnamefont
  {Stoof}},\ }\href {\doibase 10.1103/PhysRevA.47.4114} {\bibfield  {journal}
  {\bibinfo  {journal} {Phys. Rev. A}\ }\textbf {\bibinfo {volume} {47}},\
  \bibinfo {pages} {4114} (\bibinfo {year} {1993})}\BibitemShut {NoStop}%
\bibitem [{\citenamefont {Chin}\ \emph {et~al.}(2010)\citenamefont {Chin},
  \citenamefont {Grimm}, \citenamefont {Julienne},\ and\ \citenamefont
  {Tiesinga}}]{RevModPhys.82.1225}%
  \BibitemOpen
  \bibfield  {author} {\bibinfo {author} {\bibfnamefont {C.}~\bibnamefont
  {Chin}}, \bibinfo {author} {\bibfnamefont {R.}~\bibnamefont {Grimm}},
  \bibinfo {author} {\bibfnamefont {P.}~\bibnamefont {Julienne}}, \ and\
  \bibinfo {author} {\bibfnamefont {E.}~\bibnamefont {Tiesinga}},\ }\href
  {\doibase 10.1103/RevModPhys.82.1225} {\bibfield  {journal} {\bibinfo
  {journal} {Rev. Mod. Phys.}\ }\textbf {\bibinfo {volume} {82}},\ \bibinfo
  {pages} {1225} (\bibinfo {year} {2010})}\BibitemShut {NoStop}%
\bibitem [{\citenamefont {Giorgini}\ \emph {et~al.}(2008)\citenamefont
  {Giorgini}, \citenamefont {Pitaevskii},\ and\ \citenamefont
  {Stringari}}]{RevModPhys.80.1215}%
  \BibitemOpen
  \bibfield  {author} {\bibinfo {author} {\bibfnamefont {S.}~\bibnamefont
  {Giorgini}}, \bibinfo {author} {\bibfnamefont {L.~P.}\ \bibnamefont
  {Pitaevskii}}, \ and\ \bibinfo {author} {\bibfnamefont {S.}~\bibnamefont
  {Stringari}},\ }\href {\doibase 10.1103/RevModPhys.80.1215} {\bibfield
  {journal} {\bibinfo  {journal} {Rev. Mod. Phys.}\ }\textbf {\bibinfo {volume}
  {80}},\ \bibinfo {pages} {1215} (\bibinfo {year} {2008})}\BibitemShut
  {NoStop}%
\bibitem [{\citenamefont {K\"ohler}\ \emph {et~al.}(2006)\citenamefont
  {K\"ohler}, \citenamefont {G\'oral},\ and\ \citenamefont
  {Julienne}}]{RevModPhys.78.1311}%
  \BibitemOpen
  \bibfield  {author} {\bibinfo {author} {\bibfnamefont {T.}~\bibnamefont
  {K\"ohler}}, \bibinfo {author} {\bibfnamefont {K.}~\bibnamefont {G\'oral}}, \
  and\ \bibinfo {author} {\bibfnamefont {P.~S.}\ \bibnamefont {Julienne}},\
  }\href {\doibase 10.1103/RevModPhys.78.1311} {\bibfield  {journal} {\bibinfo
  {journal} {Rev. Mod. Phys.}\ }\textbf {\bibinfo {volume} {78}},\ \bibinfo
  {pages} {1311} (\bibinfo {year} {2006})}\BibitemShut {NoStop}%
\bibitem [{\citenamefont {Danzl}\ \emph {et~al.}(2010)\citenamefont {Danzl},
  \citenamefont {Mark}, \citenamefont {Haller}, \citenamefont {Gustavsson},
  \citenamefont {Hart}, \citenamefont {Aldegunde}, \citenamefont {Hutson},\
  and\ \citenamefont {N\"agerl}}]{Danzl.rovib}%
  \BibitemOpen
  \bibfield  {author} {\bibinfo {author} {\bibfnamefont {J.~G.}\ \bibnamefont
  {Danzl}}, \bibinfo {author} {\bibfnamefont {M.~J.}\ \bibnamefont {Mark}},
  \bibinfo {author} {\bibfnamefont {E.}~\bibnamefont {Haller}}, \bibinfo
  {author} {\bibfnamefont {M.}~\bibnamefont {Gustavsson}}, \bibinfo {author}
  {\bibfnamefont {R.}~\bibnamefont {Hart}}, \bibinfo {author} {\bibfnamefont
  {J.}~\bibnamefont {Aldegunde}}, \bibinfo {author} {\bibfnamefont {J.~M.}\
  \bibnamefont {Hutson}}, \ and\ \bibinfo {author} {\bibfnamefont {H.-C.}\
  \bibnamefont {N\"agerl}},\ }\href {\doibase 10.1038/nphys1533} {\bibfield
  {journal} {\bibinfo  {journal} {Nature Physics}\ }\textbf {\bibinfo {volume}
  {6}},\ \bibinfo {pages} {265 } (\bibinfo {year} {2010})}\BibitemShut
  {NoStop}%
\bibitem [{\citenamefont {Ni}\ \emph {et~al.}(2008)\citenamefont {Ni},
  \citenamefont {Ospelkaus}, \citenamefont {de~Miranda}, \citenamefont {Pe'er},
  \citenamefont {Neyenhuis}, \citenamefont {Zirbel}, \citenamefont
  {Kotochigova}, \citenamefont {Julienne}, \citenamefont {Jin},\ and\
  \citenamefont {Ye}}]{Ni10102008}%
  \BibitemOpen
  \bibfield  {author} {\bibinfo {author} {\bibfnamefont {K.-K.}\ \bibnamefont
  {Ni}}, \bibinfo {author} {\bibfnamefont {S.}~\bibnamefont {Ospelkaus}},
  \bibinfo {author} {\bibfnamefont {M.~H.~G.}\ \bibnamefont {de~Miranda}},
  \bibinfo {author} {\bibfnamefont {A.}~\bibnamefont {Pe'er}}, \bibinfo
  {author} {\bibfnamefont {B.}~\bibnamefont {Neyenhuis}}, \bibinfo {author}
  {\bibfnamefont {J.~J.}\ \bibnamefont {Zirbel}}, \bibinfo {author}
  {\bibfnamefont {S.}~\bibnamefont {Kotochigova}}, \bibinfo {author}
  {\bibfnamefont {P.~S.}\ \bibnamefont {Julienne}}, \bibinfo {author}
  {\bibfnamefont {D.~S.}\ \bibnamefont {Jin}}, \ and\ \bibinfo {author}
  {\bibfnamefont {J.}~\bibnamefont {Ye}},\ }\href {\doibase
  10.1126/science.1163861} {\bibfield  {journal} {\bibinfo  {journal}
  {Science}\ }\textbf {\bibinfo {volume} {322}},\ \bibinfo {pages} {231}
  (\bibinfo {year} {2008})}\BibitemShut {NoStop}%
\bibitem [{\citenamefont {Volz}\ \emph {et~al.}(2003)\citenamefont {Volz},
  \citenamefont {D\"urr}, \citenamefont {Ernst}, \citenamefont {Marte},\ and\
  \citenamefont {Rempe}}]{PhysRevA.68.010702}%
  \BibitemOpen
  \bibfield  {author} {\bibinfo {author} {\bibfnamefont {T.}~\bibnamefont
  {Volz}}, \bibinfo {author} {\bibfnamefont {S.}~\bibnamefont {D\"urr}},
  \bibinfo {author} {\bibfnamefont {S.}~\bibnamefont {Ernst}}, \bibinfo
  {author} {\bibfnamefont {A.}~\bibnamefont {Marte}}, \ and\ \bibinfo {author}
  {\bibfnamefont {G.}~\bibnamefont {Rempe}},\ }\href {\doibase
  10.1103/PhysRevA.68.010702} {\bibfield  {journal} {\bibinfo  {journal} {Phys.
  Rev. A}\ }\textbf {\bibinfo {volume} {68}},\ \bibinfo {pages} {010702}
  (\bibinfo {year} {2003})}\BibitemShut {NoStop}%
\bibitem [{\citenamefont {Roberts}\ \emph {et~al.}(1998)\citenamefont
  {Roberts}, \citenamefont {Claussen}, \citenamefont {Burke}, \citenamefont
  {Greene}, \citenamefont {Cornell},\ and\ \citenamefont
  {Wieman}}]{PhysRevLett.81.5109}%
  \BibitemOpen
  \bibfield  {author} {\bibinfo {author} {\bibfnamefont {J.~L.}\ \bibnamefont
  {Roberts}}, \bibinfo {author} {\bibfnamefont {N.~R.}\ \bibnamefont
  {Claussen}}, \bibinfo {author} {\bibfnamefont {J.~P.}\ \bibnamefont {Burke}},
  \bibinfo {author} {\bibfnamefont {C.~H.}\ \bibnamefont {Greene}}, \bibinfo
  {author} {\bibfnamefont {E.~A.}\ \bibnamefont {Cornell}}, \ and\ \bibinfo
  {author} {\bibfnamefont {C.~E.}\ \bibnamefont {Wieman}},\ }\href {\doibase
  10.1103/PhysRevLett.81.5109} {\bibfield  {journal} {\bibinfo  {journal}
  {Phys. Rev. Lett.}\ }\textbf {\bibinfo {volume} {81}},\ \bibinfo {pages}
  {5109} (\bibinfo {year} {1998})}\BibitemShut {NoStop}%
\bibitem [{\citenamefont {Claussen}\ \emph {et~al.}(2003)\citenamefont
  {Claussen}, \citenamefont {Kokkelmans}, \citenamefont {Thompson},
  \citenamefont {Donley}, \citenamefont {Hodby},\ and\ \citenamefont
  {Wieman}}]{PhysRevA.67.060701}%
  \BibitemOpen
  \bibfield  {author} {\bibinfo {author} {\bibfnamefont {N.~R.}\ \bibnamefont
  {Claussen}}, \bibinfo {author} {\bibfnamefont {S.~J. J. M.~F.}\ \bibnamefont
  {Kokkelmans}}, \bibinfo {author} {\bibfnamefont {S.~T.}\ \bibnamefont
  {Thompson}}, \bibinfo {author} {\bibfnamefont {E.~A.}\ \bibnamefont
  {Donley}}, \bibinfo {author} {\bibfnamefont {E.}~\bibnamefont {Hodby}}, \
  and\ \bibinfo {author} {\bibfnamefont {C.~E.}\ \bibnamefont {Wieman}},\
  }\href {\doibase 10.1103/PhysRevA.67.060701} {\bibfield  {journal} {\bibinfo
  {journal} {Phys. Rev. A}\ }\textbf {\bibinfo {volume} {67}},\ \bibinfo
  {pages} {060701} (\bibinfo {year} {2003})}\BibitemShut {NoStop}%
\bibitem [{\citenamefont {Cornish}\ \emph {et~al.}(2000)\citenamefont
  {Cornish}, \citenamefont {Claussen}, \citenamefont {Roberts}, \citenamefont
  {Cornell},\ and\ \citenamefont {Wieman}}]{PhysRevLett.85.1795}%
  \BibitemOpen
  \bibfield  {author} {\bibinfo {author} {\bibfnamefont {S.~L.}\ \bibnamefont
  {Cornish}}, \bibinfo {author} {\bibfnamefont {N.~R.}\ \bibnamefont
  {Claussen}}, \bibinfo {author} {\bibfnamefont {J.~L.}\ \bibnamefont
  {Roberts}}, \bibinfo {author} {\bibfnamefont {E.~A.}\ \bibnamefont
  {Cornell}}, \ and\ \bibinfo {author} {\bibfnamefont {C.~E.}\ \bibnamefont
  {Wieman}},\ }\href {\doibase 10.1103/PhysRevLett.85.1795} {\bibfield
  {journal} {\bibinfo  {journal} {Phys. Rev. Lett.}\ }\textbf {\bibinfo
  {volume} {85}},\ \bibinfo {pages} {1795} (\bibinfo {year}
  {2000})}\BibitemShut {NoStop}%
\bibitem [{\citenamefont {Roberts}\ \emph
  {et~al.}(2001{\natexlab{a}})\citenamefont {Roberts}, \citenamefont
  {Claussen}, \citenamefont {Cornish}, \citenamefont {Donley}, \citenamefont
  {Cornell},\ and\ \citenamefont {Wieman}}]{PhysRevLett.86.4211}%
  \BibitemOpen
  \bibfield  {author} {\bibinfo {author} {\bibfnamefont {J.~L.}\ \bibnamefont
  {Roberts}}, \bibinfo {author} {\bibfnamefont {N.~R.}\ \bibnamefont
  {Claussen}}, \bibinfo {author} {\bibfnamefont {S.~L.}\ \bibnamefont
  {Cornish}}, \bibinfo {author} {\bibfnamefont {E.~A.}\ \bibnamefont {Donley}},
  \bibinfo {author} {\bibfnamefont {E.~A.}\ \bibnamefont {Cornell}}, \ and\
  \bibinfo {author} {\bibfnamefont {C.~E.}\ \bibnamefont {Wieman}},\ }\href
  {\doibase 10.1103/PhysRevLett.86.4211} {\bibfield  {journal} {\bibinfo
  {journal} {Phys. Rev. Lett.}\ }\textbf {\bibinfo {volume} {86}},\ \bibinfo
  {pages} {4211} (\bibinfo {year} {2001}{\natexlab{a}})}\BibitemShut {NoStop}%
\bibitem [{\citenamefont {{Donley}}\ \emph {et~al.}(2001)\citenamefont
  {{Donley}}, \citenamefont {{Claussen}}, \citenamefont {{Cornish}},
  \citenamefont {{Roberts}}, \citenamefont {{Cornell}},\ and\ \citenamefont
  {{Wieman}}}]{Nature.Collapse}%
  \BibitemOpen
  \bibfield  {author} {\bibinfo {author} {\bibfnamefont {E.~A.}\ \bibnamefont
  {{Donley}}}, \bibinfo {author} {\bibfnamefont {N.~R.}\ \bibnamefont
  {{Claussen}}}, \bibinfo {author} {\bibfnamefont {S.~L.}\ \bibnamefont
  {{Cornish}}}, \bibinfo {author} {\bibfnamefont {J.~L.}\ \bibnamefont
  {{Roberts}}}, \bibinfo {author} {\bibfnamefont {E.~A.}\ \bibnamefont
  {{Cornell}}}, \ and\ \bibinfo {author} {\bibfnamefont {C.~E.}\ \bibnamefont
  {{Wieman}}},\ }\href {\doibase 10.1038/35085500} {\bibfield  {journal}
  {\bibinfo  {journal} {\nat}\ }\textbf {\bibinfo {volume} {412}},\ \bibinfo
  {pages} {295} (\bibinfo {year} {2001})}\BibitemShut {NoStop}%
\bibitem [{\citenamefont {Cornish}\ \emph {et~al.}(2006)\citenamefont
  {Cornish}, \citenamefont {Thompson},\ and\ \citenamefont
  {Wieman}}]{PhysRevLett.96.170401}%
  \BibitemOpen
  \bibfield  {author} {\bibinfo {author} {\bibfnamefont {S.~L.}\ \bibnamefont
  {Cornish}}, \bibinfo {author} {\bibfnamefont {S.~T.}\ \bibnamefont
  {Thompson}}, \ and\ \bibinfo {author} {\bibfnamefont {C.~E.}\ \bibnamefont
  {Wieman}},\ }\href {\doibase 10.1103/PhysRevLett.96.170401} {\bibfield
  {journal} {\bibinfo  {journal} {Phys. Rev. Lett.}\ }\textbf {\bibinfo
  {volume} {96}},\ \bibinfo {pages} {170401} (\bibinfo {year}
  {2006})}\BibitemShut {NoStop}%
\bibitem [{\citenamefont {Papp}\ \emph
  {et~al.}(2008{\natexlab{a}})\citenamefont {Papp}, \citenamefont {Pino},
  \citenamefont {Wild}, \citenamefont {Ronen}, \citenamefont {Wieman},
  \citenamefont {Jin},\ and\ \citenamefont {Cornell}}]{PhysRevLett.101.135301}%
  \BibitemOpen
  \bibfield  {author} {\bibinfo {author} {\bibfnamefont {S.~B.}\ \bibnamefont
  {Papp}}, \bibinfo {author} {\bibfnamefont {J.~M.}\ \bibnamefont {Pino}},
  \bibinfo {author} {\bibfnamefont {R.~J.}\ \bibnamefont {Wild}}, \bibinfo
  {author} {\bibfnamefont {S.}~\bibnamefont {Ronen}}, \bibinfo {author}
  {\bibfnamefont {C.~E.}\ \bibnamefont {Wieman}}, \bibinfo {author}
  {\bibfnamefont {D.~S.}\ \bibnamefont {Jin}}, \ and\ \bibinfo {author}
  {\bibfnamefont {E.~A.}\ \bibnamefont {Cornell}},\ }\href {\doibase
  10.1103/PhysRevLett.101.135301} {\bibfield  {journal} {\bibinfo  {journal}
  {Phys. Rev. Lett.}\ }\textbf {\bibinfo {volume} {101}},\ \bibinfo {pages}
  {135301} (\bibinfo {year} {2008}{\natexlab{a}})}\BibitemShut {NoStop}%
\bibitem [{\citenamefont {Altin}\ \emph {et~al.}(2011)\citenamefont {Altin},
  \citenamefont {Dennis}, \citenamefont {McDonald}, \citenamefont {D\"oring},
  \citenamefont {Debs}, \citenamefont {Close}, \citenamefont {Savage},\ and\
  \citenamefont {Robins}}]{PhysRevA.84.033632}%
  \BibitemOpen
  \bibfield  {author} {\bibinfo {author} {\bibfnamefont {P.~A.}\ \bibnamefont
  {Altin}}, \bibinfo {author} {\bibfnamefont {G.~R.}\ \bibnamefont {Dennis}},
  \bibinfo {author} {\bibfnamefont {G.~D.}\ \bibnamefont {McDonald}}, \bibinfo
  {author} {\bibfnamefont {D.}~\bibnamefont {D\"oring}}, \bibinfo {author}
  {\bibfnamefont {J.~E.}\ \bibnamefont {Debs}}, \bibinfo {author}
  {\bibfnamefont {J.~D.}\ \bibnamefont {Close}}, \bibinfo {author}
  {\bibfnamefont {C.~M.}\ \bibnamefont {Savage}}, \ and\ \bibinfo {author}
  {\bibfnamefont {N.~P.}\ \bibnamefont {Robins}},\ }\href {\doibase
  10.1103/PhysRevA.84.033632} {\bibfield  {journal} {\bibinfo  {journal} {Phys.
  Rev. A}\ }\textbf {\bibinfo {volume} {84}},\ \bibinfo {pages} {033632}
  (\bibinfo {year} {2011})}\BibitemShut {NoStop}%
\bibitem [{\citenamefont {Burke}\ \emph {et~al.}(1998)\citenamefont {Burke},
  \citenamefont {Bohn}, \citenamefont {Esry},\ and\ \citenamefont
  {Greene}}]{PhysRevLett.80.2097}%
  \BibitemOpen
  \bibfield  {author} {\bibinfo {author} {\bibfnamefont {J.~P.}\ \bibnamefont
  {Burke}}, \bibinfo {author} {\bibfnamefont {J.~L.}\ \bibnamefont {Bohn}},
  \bibinfo {author} {\bibfnamefont {B.~D.}\ \bibnamefont {Esry}}, \ and\
  \bibinfo {author} {\bibfnamefont {C.~H.}\ \bibnamefont {Greene}},\ }\href
  {\doibase 10.1103/PhysRevLett.80.2097} {\bibfield  {journal} {\bibinfo
  {journal} {Phys. Rev. Lett.}\ }\textbf {\bibinfo {volume} {80}},\ \bibinfo
  {pages} {2097} (\bibinfo {year} {1998})}\BibitemShut {NoStop}%
\bibitem [{\citenamefont {Roberts}\ \emph {et~al.}(2000)\citenamefont
  {Roberts}, \citenamefont {Claussen}, \citenamefont {Cornish},\ and\
  \citenamefont {Wieman}}]{PhysRevLett.85.728}%
  \BibitemOpen
  \bibfield  {author} {\bibinfo {author} {\bibfnamefont {J.~L.}\ \bibnamefont
  {Roberts}}, \bibinfo {author} {\bibfnamefont {N.~R.}\ \bibnamefont
  {Claussen}}, \bibinfo {author} {\bibfnamefont {S.~L.}\ \bibnamefont
  {Cornish}}, \ and\ \bibinfo {author} {\bibfnamefont {C.~E.}\ \bibnamefont
  {Wieman}},\ }\href {\doibase 10.1103/PhysRevLett.85.728} {\bibfield
  {journal} {\bibinfo  {journal} {Phys. Rev. Lett.}\ }\textbf {\bibinfo
  {volume} {85}},\ \bibinfo {pages} {728} (\bibinfo {year} {2000})}\BibitemShut
  {NoStop}%
\bibitem [{\citenamefont {Papp}\ \emph
  {et~al.}(2008{\natexlab{b}})\citenamefont {Papp}, \citenamefont {Pino},\ and\
  \citenamefont {Wieman}}]{PhysRevLett.101.040402}%
  \BibitemOpen
  \bibfield  {author} {\bibinfo {author} {\bibfnamefont {S.~B.}\ \bibnamefont
  {Papp}}, \bibinfo {author} {\bibfnamefont {J.~M.}\ \bibnamefont {Pino}}, \
  and\ \bibinfo {author} {\bibfnamefont {C.~E.}\ \bibnamefont {Wieman}},\
  }\href {\doibase 10.1103/PhysRevLett.101.040402} {\bibfield  {journal}
  {\bibinfo  {journal} {Phys. Rev. Lett.}\ }\textbf {\bibinfo {volume} {101}},\
  \bibinfo {pages} {040402} (\bibinfo {year} {2008}{\natexlab{b}})}\BibitemShut
  {NoStop}%
\bibitem [{\citenamefont {Altin}\ \emph {et~al.}(2010)\citenamefont {Altin},
  \citenamefont {Robins}, \citenamefont {D\"oring}, \citenamefont {Debs},
  \citenamefont {Poldy}, \citenamefont {Figl},\ and\ \citenamefont
  {Close}}]{Altin.RevSciInst}%
  \BibitemOpen
  \bibfield  {author} {\bibinfo {author} {\bibfnamefont {P.~A.}\ \bibnamefont
  {Altin}}, \bibinfo {author} {\bibfnamefont {N.~P.}\ \bibnamefont {Robins}},
  \bibinfo {author} {\bibfnamefont {D.}~\bibnamefont {D\"oring}}, \bibinfo
  {author} {\bibfnamefont {J.~E.}\ \bibnamefont {Debs}}, \bibinfo {author}
  {\bibfnamefont {R.}~\bibnamefont {Poldy}}, \bibinfo {author} {\bibfnamefont
  {C.}~\bibnamefont {Figl}}, \ and\ \bibinfo {author} {\bibfnamefont {J.~D.}\
  \bibnamefont {Close}},\ }\href@noop {} {\bibfield  {journal} {\bibinfo
  {journal} {Rev. Sci. Instrum.}\ }\textbf {\bibinfo {volume} {81}},\ \bibinfo
  {pages} {063103} (\bibinfo {year} {2010})}\BibitemShut {NoStop}%
\bibitem [{\citenamefont {H\"andel}\ \emph {et~al.}(2012)\citenamefont
  {H\"andel}, \citenamefont {Marchant}, \citenamefont {Wiles}, \citenamefont
  {Hopkins},\ and\ \citenamefont {Cornish}}]{SylviApparatus}%
  \BibitemOpen
  \bibfield  {author} {\bibinfo {author} {\bibfnamefont {S.}~\bibnamefont
  {H\"andel}}, \bibinfo {author} {\bibfnamefont {A.~L.}\ \bibnamefont
  {Marchant}}, \bibinfo {author} {\bibfnamefont {T.~P.}\ \bibnamefont {Wiles}},
  \bibinfo {author} {\bibfnamefont {S.~A.}\ \bibnamefont {Hopkins}}, \ and\
  \bibinfo {author} {\bibfnamefont {S.~L.}\ \bibnamefont {Cornish}},\
  }\href@noop {} {\bibfield  {journal} {\bibinfo  {journal} {Rev. Sci.
  Instrum.}\ }\textbf {\bibinfo {volume} {83}},\ \bibinfo {pages} {013105}
  (\bibinfo {year} {2012})}\BibitemShut {NoStop}%
\bibitem [{\citenamefont {Ketterle}\ and\ \citenamefont {van
  Druten}(1996)}]{KetterleReview}%
  \BibitemOpen
  \bibfield  {author} {\bibinfo {author} {\bibfnamefont {W.}~\bibnamefont
  {Ketterle}}\ and\ \bibinfo {author} {\bibfnamefont {N.}~\bibnamefont {van
  Druten}},\ }\href {\doibase 10.1016/S1049-250X(08)60101-9} {\bibfield
  {journal} {\bibinfo  {journal} {Adv. Atom. Mol. Opt. Phys}\ }\textbf
  {\bibinfo {volume} {37}},\ \bibinfo {pages} {181} (\bibinfo {year}
  {1996})}\BibitemShut {NoStop}%
\bibitem [{\citenamefont {Lin}\ \emph {et~al.}(2009)\citenamefont {Lin},
  \citenamefont {Perry}, \citenamefont {Compton}, \citenamefont {Spielman},\
  and\ \citenamefont {Porto}}]{PhysRevA.79.063631}%
  \BibitemOpen
  \bibfield  {author} {\bibinfo {author} {\bibfnamefont {Y.-J.}\ \bibnamefont
  {Lin}}, \bibinfo {author} {\bibfnamefont {A.~R.}\ \bibnamefont {Perry}},
  \bibinfo {author} {\bibfnamefont {R.~L.}\ \bibnamefont {Compton}}, \bibinfo
  {author} {\bibfnamefont {I.~B.}\ \bibnamefont {Spielman}}, \ and\ \bibinfo
  {author} {\bibfnamefont {J.~V.}\ \bibnamefont {Porto}},\ }\href {\doibase
  10.1103/PhysRevA.79.063631} {\bibfield  {journal} {\bibinfo  {journal} {Phys.
  Rev. A}\ }\textbf {\bibinfo {volume} {79}},\ \bibinfo {pages} {063631}
  (\bibinfo {year} {2009})}\BibitemShut {NoStop}%
\bibitem [{\citenamefont {Julienne}()}]{PJprivate}%
  \BibitemOpen
  \bibfield  {author} {\bibinfo {author} {\bibfnamefont {P.}~\bibnamefont
  {Julienne}},\ }\href@noop {} {}\bibinfo {howpublished} {Private
  communication}\BibitemShut {NoStop}%
\bibitem [{\citenamefont {Weber}\ \emph {et~al.}(2003)\citenamefont {Weber},
  \citenamefont {Herbig}, \citenamefont {Mark}, \citenamefont {N\"agerl},\ and\
  \citenamefont {Grimm}}]{PhysRevLett.91.123201}%
  \BibitemOpen
  \bibfield  {author} {\bibinfo {author} {\bibfnamefont {T.}~\bibnamefont
  {Weber}}, \bibinfo {author} {\bibfnamefont {J.}~\bibnamefont {Herbig}},
  \bibinfo {author} {\bibfnamefont {M.}~\bibnamefont {Mark}}, \bibinfo {author}
  {\bibfnamefont {H.-C.}\ \bibnamefont {N\"agerl}}, \ and\ \bibinfo {author}
  {\bibfnamefont {R.}~\bibnamefont {Grimm}},\ }\href {\doibase
  10.1103/PhysRevLett.91.123201} {\bibfield  {journal} {\bibinfo  {journal}
  {Phys. Rev. Lett.}\ }\textbf {\bibinfo {volume} {91}},\ \bibinfo {pages}
  {123201} (\bibinfo {year} {2003})}\BibitemShut {NoStop}%
\bibitem [{\citenamefont {Esry}\ \emph {et~al.}(1999)\citenamefont {Esry},
  \citenamefont {Greene},\ and\ \citenamefont {Burke}}]{PhysRevLett.83.1751}%
  \BibitemOpen
  \bibfield  {author} {\bibinfo {author} {\bibfnamefont {B.~D.}\ \bibnamefont
  {Esry}}, \bibinfo {author} {\bibfnamefont {C.~H.}\ \bibnamefont {Greene}}, \
  and\ \bibinfo {author} {\bibfnamefont {J.~P.}\ \bibnamefont {Burke}},\ }\href
  {\doibase 10.1103/PhysRevLett.83.1751} {\bibfield  {journal} {\bibinfo
  {journal} {Phys. Rev. Lett.}\ }\textbf {\bibinfo {volume} {83}},\ \bibinfo
  {pages} {1751} (\bibinfo {year} {1999})}\BibitemShut {NoStop}%
\bibitem [{\citenamefont {Ruprecht}\ \emph {et~al.}(1995)\citenamefont
  {Ruprecht}, \citenamefont {Holland}, \citenamefont {Burnett},\ and\
  \citenamefont {Edwards}}]{PhysRevA.51.4704}%
  \BibitemOpen
  \bibfield  {author} {\bibinfo {author} {\bibfnamefont {P.~A.}\ \bibnamefont
  {Ruprecht}}, \bibinfo {author} {\bibfnamefont {M.~J.}\ \bibnamefont
  {Holland}}, \bibinfo {author} {\bibfnamefont {K.}~\bibnamefont {Burnett}}, \
  and\ \bibinfo {author} {\bibfnamefont {M.}~\bibnamefont {Edwards}},\ }\href
  {\doibase 10.1103/PhysRevA.51.4704} {\bibfield  {journal} {\bibinfo
  {journal} {Phys. Rev. A}\ }\textbf {\bibinfo {volume} {51}},\ \bibinfo
  {pages} {4704} (\bibinfo {year} {1995})}\BibitemShut {NoStop}%
\bibitem [{\citenamefont {Roberts}\ \emph
  {et~al.}(2001{\natexlab{b}})\citenamefont {Roberts}, \citenamefont {Burke},
  \citenamefont {Claussen}, \citenamefont {Cornish}, \citenamefont {Donley},\
  and\ \citenamefont {Wieman}}]{PhysRevA.64.024702}%
  \BibitemOpen
  \bibfield  {author} {\bibinfo {author} {\bibfnamefont {J.~L.}\ \bibnamefont
  {Roberts}}, \bibinfo {author} {\bibfnamefont {J.~P.}\ \bibnamefont {Burke}},
  \bibinfo {author} {\bibfnamefont {N.~R.}\ \bibnamefont {Claussen}}, \bibinfo
  {author} {\bibfnamefont {S.~L.}\ \bibnamefont {Cornish}}, \bibinfo {author}
  {\bibfnamefont {E.~A.}\ \bibnamefont {Donley}}, \ and\ \bibinfo {author}
  {\bibfnamefont {C.~E.}\ \bibnamefont {Wieman}},\ }\href {\doibase
  10.1103/PhysRevA.64.024702} {\bibfield  {journal} {\bibinfo  {journal} {Phys.
  Rev. A}\ }\textbf {\bibinfo {volume} {64}},\ \bibinfo {pages} {024702}
  (\bibinfo {year} {2001}{\natexlab{b}})}\BibitemShut {NoStop}%
\bibitem [{\citenamefont {Dalfovo}\ \emph {et~al.}(1999)\citenamefont
  {Dalfovo}, \citenamefont {Giorgini}, \citenamefont {Pitaevskii},\ and\
  \citenamefont {Stringari}}]{RevModPhys.71.463}%
  \BibitemOpen
  \bibfield  {author} {\bibinfo {author} {\bibfnamefont {F.}~\bibnamefont
  {Dalfovo}}, \bibinfo {author} {\bibfnamefont {S.}~\bibnamefont {Giorgini}},
  \bibinfo {author} {\bibfnamefont {L.~P.}\ \bibnamefont {Pitaevskii}}, \ and\
  \bibinfo {author} {\bibfnamefont {S.}~\bibnamefont {Stringari}},\ }\href
  {\doibase 10.1103/RevModPhys.71.463} {\bibfield  {journal} {\bibinfo
  {journal} {Rev. Mod. Phys.}\ }\textbf {\bibinfo {volume} {71}},\ \bibinfo
  {pages} {463} (\bibinfo {year} {1999})}\BibitemShut {NoStop}%
\bibitem [{\citenamefont {McCarron}\ \emph {et~al.}(2011)\citenamefont
  {McCarron}, \citenamefont {Cho}, \citenamefont {Jenkin}, \citenamefont
  {K\"oppinger},\ and\ \citenamefont {Cornish}}]{PhysRevA.84.011603}%
  \BibitemOpen
  \bibfield  {author} {\bibinfo {author} {\bibfnamefont {D.~J.}\ \bibnamefont
  {McCarron}}, \bibinfo {author} {\bibfnamefont {H.~W.}\ \bibnamefont {Cho}},
  \bibinfo {author} {\bibfnamefont {D.~L.}\ \bibnamefont {Jenkin}}, \bibinfo
  {author} {\bibfnamefont {M.~P.}\ \bibnamefont {K\"oppinger}}, \ and\ \bibinfo
  {author} {\bibfnamefont {S.~L.}\ \bibnamefont {Cornish}},\ }\href {\doibase
  10.1103/PhysRevA.84.011603} {\bibfield  {journal} {\bibinfo  {journal} {Phys.
  Rev. A}\ }\textbf {\bibinfo {volume} {84}},\ \bibinfo {pages} {011603}
  (\bibinfo {year} {2011})}\BibitemShut {NoStop}%
\bibitem [{\citenamefont {{Strecker}}\ \emph {et~al.}(2002)\citenamefont
  {{Strecker}}, \citenamefont {{Partridge}}, \citenamefont {{Truscott}},\ and\
  \citenamefont {{Hulet}}}]{2002Natur.417..150S}%
  \BibitemOpen
  \bibfield  {author} {\bibinfo {author} {\bibfnamefont {K.~E.}\ \bibnamefont
  {{Strecker}}}, \bibinfo {author} {\bibfnamefont {G.~B.}\ \bibnamefont
  {{Partridge}}}, \bibinfo {author} {\bibfnamefont {A.~G.}\ \bibnamefont
  {{Truscott}}}, \ and\ \bibinfo {author} {\bibfnamefont {R.~G.}\ \bibnamefont
  {{Hulet}}},\ }\href {\doibase 10.1038/nature747} {\bibfield  {journal}
  {\bibinfo  {journal} {\nat}\ }\textbf {\bibinfo {volume} {417}},\ \bibinfo
  {pages} {150} (\bibinfo {year} {2002})}\BibitemShut {NoStop}%
\bibitem [{\citenamefont {{Khaykovich}}\ \emph {et~al.}(2002)\citenamefont
  {{Khaykovich}}, \citenamefont {{Schreck}}, \citenamefont {{Ferrari}},
  \citenamefont {{Bourdel}}, \citenamefont {{Cubizolles}}, \citenamefont
  {{Carr}}, \citenamefont {{Castin}},\ and\ \citenamefont
  {{Salomon}}}]{2002Sci...296.1290K}%
  \BibitemOpen
  \bibfield  {author} {\bibinfo {author} {\bibfnamefont {L.}~\bibnamefont
  {{Khaykovich}}}, \bibinfo {author} {\bibfnamefont {F.}~\bibnamefont
  {{Schreck}}}, \bibinfo {author} {\bibfnamefont {G.}~\bibnamefont
  {{Ferrari}}}, \bibinfo {author} {\bibfnamefont {T.}~\bibnamefont
  {{Bourdel}}}, \bibinfo {author} {\bibfnamefont {J.}~\bibnamefont
  {{Cubizolles}}}, \bibinfo {author} {\bibfnamefont {L.~D.}\ \bibnamefont
  {{Carr}}}, \bibinfo {author} {\bibfnamefont {Y.}~\bibnamefont {{Castin}}}, \
  and\ \bibinfo {author} {\bibfnamefont {C.}~\bibnamefont {{Salomon}}},\ }\href
  {\doibase 10.1126/science.1071021} {\bibfield  {journal} {\bibinfo  {journal}
  {Science}\ }\textbf {\bibinfo {volume} {296}},\ \bibinfo {pages} {1290}
  (\bibinfo {year} {2002})}\BibitemShut {NoStop}%
\bibitem [{\citenamefont {Cornish}\ \emph {et~al.}(2009)\citenamefont
  {Cornish}, \citenamefont {Parker}, \citenamefont {Martin}, \citenamefont
  {Judd}, \citenamefont {Scott}, \citenamefont {Fromhold},\ and\ \citenamefont
  {Adams}}]{Cornish20091299}%
  \BibitemOpen
  \bibfield  {author} {\bibinfo {author} {\bibfnamefont {S.}~\bibnamefont
  {Cornish}}, \bibinfo {author} {\bibfnamefont {N.}~\bibnamefont {Parker}},
  \bibinfo {author} {\bibfnamefont {A.}~\bibnamefont {Martin}}, \bibinfo
  {author} {\bibfnamefont {T.}~\bibnamefont {Judd}}, \bibinfo {author}
  {\bibfnamefont {R.}~\bibnamefont {Scott}}, \bibinfo {author} {\bibfnamefont
  {T.}~\bibnamefont {Fromhold}}, \ and\ \bibinfo {author} {\bibfnamefont
  {C.}~\bibnamefont {Adams}},\ }\href {\doibase 10.1016/j.physd.2008.07.011}
  {\bibfield  {journal} {\bibinfo  {journal} {Physica D: Nonlinear Phenomena}\
  }\textbf {\bibinfo {volume} {238}},\ \bibinfo {pages} {1299 } (\bibinfo
  {year} {2009})}\BibitemShut {NoStop}%
\bibitem [{\citenamefont {Parker}\ \emph {et~al.}(2008)\citenamefont {Parker},
  \citenamefont {Martin}, \citenamefont {Cornish},\ and\ \citenamefont
  {Adams}}]{0953-4075-41-4-045303}%
  \BibitemOpen
  \bibfield  {author} {\bibinfo {author} {\bibfnamefont {N.~G.}\ \bibnamefont
  {Parker}}, \bibinfo {author} {\bibfnamefont {A.~M.}\ \bibnamefont {Martin}},
  \bibinfo {author} {\bibfnamefont {S.~L.}\ \bibnamefont {Cornish}}, \ and\
  \bibinfo {author} {\bibfnamefont {C.~S.}\ \bibnamefont {Adams}},\ }\href@noop
  {} {\bibfield  {journal} {\bibinfo  {journal} {J. Phys. B}\ }\textbf
  {\bibinfo {volume} {41}},\ \bibinfo {pages} {045303} (\bibinfo {year}
  {2008})}\BibitemShut {NoStop}%
\bibitem [{\citenamefont {Billam}\ \emph {et~al.}(2011)\citenamefont {Billam},
  \citenamefont {Cornish},\ and\ \citenamefont
  {Gardiner}}]{PhysRevA.83.041602}%
  \BibitemOpen
  \bibfield  {author} {\bibinfo {author} {\bibfnamefont {T.~P.}\ \bibnamefont
  {Billam}}, \bibinfo {author} {\bibfnamefont {S.~L.}\ \bibnamefont {Cornish}},
  \ and\ \bibinfo {author} {\bibfnamefont {S.~A.}\ \bibnamefont {Gardiner}},\
  }\href {\doibase 10.1103/PhysRevA.83.041602} {\bibfield  {journal} {\bibinfo
  {journal} {Phys. Rev. A}\ }\textbf {\bibinfo {volume} {83}},\ \bibinfo
  {pages} {041602} (\bibinfo {year} {2011})}\BibitemShut {NoStop}%
\bibitem [{\citenamefont {Weiss}\ and\ \citenamefont
  {Castin}(2009)}]{PhysRevLett.102.010403}%
  \BibitemOpen
  \bibfield  {author} {\bibinfo {author} {\bibfnamefont {C.}~\bibnamefont
  {Weiss}}\ and\ \bibinfo {author} {\bibfnamefont {Y.}~\bibnamefont {Castin}},\
  }\href {\doibase 10.1103/PhysRevLett.102.010403} {\bibfield  {journal}
  {\bibinfo  {journal} {Phys. Rev. Lett.}\ }\textbf {\bibinfo {volume} {102}},\
  \bibinfo {pages} {010403} (\bibinfo {year} {2009})}\BibitemShut {NoStop}%
\bibitem [{\citenamefont {Streltsov}\ \emph {et~al.}(2009)\citenamefont
  {Streltsov}, \citenamefont {Alon},\ and\ \citenamefont
  {Cederbaum}}]{PhysRevA.80.043616}%
  \BibitemOpen
  \bibfield  {author} {\bibinfo {author} {\bibfnamefont {A.~I.}\ \bibnamefont
  {Streltsov}}, \bibinfo {author} {\bibfnamefont {O.~E.}\ \bibnamefont {Alon}},
  \ and\ \bibinfo {author} {\bibfnamefont {L.~S.}\ \bibnamefont {Cederbaum}},\
  }\href {\doibase 10.1103/PhysRevA.80.043616} {\bibfield  {journal} {\bibinfo
  {journal} {Phys. Rev. A}\ }\textbf {\bibinfo {volume} {80}},\ \bibinfo
  {pages} {043616} (\bibinfo {year} {2009})}\BibitemShut {NoStop}%
\end{thebibliography}

%

\end{document}